\documentclass[english,aps,prl,notitlepage,twocolumn,superscriptaddress,longbibliography]{revtex4-1}
\usepackage[T1]{fontenc}
\usepackage[latin9]{inputenc}
\usepackage{mathrsfs}
\usepackage{amsmath}
\usepackage{graphicx}

\makeatletter
\usepackage[colorlinks,citecolor=blue,linkcolor=blue]{hyperref}

\makeatother

\usepackage{babel}
\begin{document}
\title{Extrapolating the thermodynamic length with finite-time measurements}
\author{Jin-Fu Chen}
\address{Beijing Computational Science Research Center, Beijing 100193, China}
\address{Graduate School of China Academy of Engineering Physics, No. 10 Xibeiwang
East Road, Haidian District, Beijing, 100193, China}
\author{C. P. Sun}
\email{cpsun@gscaep.ac.cn}

\address{Beijing Computational Science Research Center, Beijing 100193, China}
\address{Graduate School of China Academy of Engineering Physics, No. 10 Xibeiwang
East Road, Haidian District, Beijing, 100193, China}
\author{Hui Dong}
\email{hdong@gscaep.ac.cn}

\address{Graduate School of China Academy of Engineering Physics, No. 10 Xibeiwang
East Road, Haidian District, Beijing, 100193, China}
\date{\today}
\begin{abstract}
The excess work performed in a heat-engine process with given finite
operation time $\tau$ is bounded by the thermodynamic length, which
measures the distance during the relaxation along a path in the space
of the thermodynamic state. Unfortunately, the thermodynamic length,
as a guidance for the heat engine optimization, is beyond the experimental
measurement. We propose to measure the thermodynamic length $\mathcal{L}$
through the extrapolation of finite-time measurements $\mathcal{L}(\tau)=\int_{0}^{\tau}[P_{\mathrm{ex}}(t)]^{1/2}dt$
via the excess power $P_{\mathrm{ex}}(t)$. The current proposal allows
to measure the thermodynamic length for a single control parameter
without requiring extra effort to find the optimal control scheme.
We illustrate the measurement strategy via examples of the quantum
harmonic oscillator with tuning frequency and the classical ideal
gas with changing volume.
\end{abstract}
\maketitle
\textit{Introduction.}---Optimization of thermodynamic processes
is of practical importance in finite-time thermodynamics \citep{Berry2000,Koning2005,Schmiedl2007,Andresen2011,Cavina2018,Solon2018}
to minimize the dissipation in finite-time transformation processes
\citep{Aurell2011,Aurell2012,Zulkowski2014,Proesmans2020b,Proesmans2020c},
and to improve the performance of real heat engines \citep{Curzon1975,Esposito2010,Esposito2010a,Tomas2012,Cavina2018a,Abiuso2020,Lee2020,Singh2020,Zhang2020a}.
The extent, to which the optimization can be achieved, is discovered
to be limited by the geometry properties of the thermodynamic equilibrium
state space \citep{Weinhold1975,Ruppeiner1979,Gilmore1984,Salamon1985,Ruppeiner1995,Crooks2007,Feng2009}
and provides a criterion for the amelioration of specific control
schemes. For a finite-time isothermal process with operation time
$\tau$, such optimization is realized by minimizing the entropy production
\citep{Salamon1980a,Diosi1996,Zulkowski2015,Zulkowski2015a} or the
excess work $W_{\mathrm{ex}}=W-\Delta F$ \citep{Bonan_a_2014,Bonanca2018,Zhang2020b},
where $W$ is the total performed work and $\Delta F$ is the free
energy change. For a given control scheme, the excess work at the
long-time limit is bounded as $W_{\mathrm{ex}}\geq\mathcal{L}^{2}/\tau$
\citep{Salamon1983,Nulton1985,Tu2012,Gong2016,Cavina2017,Ma2018},
where $\mathcal{L}$ is the thermodynamic length during the relaxation
dynamics \citep{Sivak2012,Zulkowski2012,Rotskoff2015,Sivak2016}.
In the recent development of finite-time quantum thermodynamics \citep{Esposito_2009,Campisi2011,Kosloff2013,Vinjanampathy2016,Strasberg2017},
the thermodynamic length relates to the geometric measures of quantum
states, such as the Bures length and the Fisher or the Wigner-Yanase
skew information \citep{Deffner2010,Deffner2013,Mancino2018,Pires2016,Bengtsson2017},
and provides a lower bound of the excess work and the entropy production
in a finite-time quantum thermodynamic process \citep{Scandi2018,Brandner2020,Abiuso2020a,Miller2020}.

However, it remains a challenge to practically measure the thermodynamics
length in a real thermodynamic system. To obtain the metric of the
state space, one needs the exact equilibrium thermal states along
the path of the control parameters, which could be the frequency of
harmonic oscillator or the volume of the classical ideal gas. One
proposal for the measurement is the utilization of the discrete-step
process \citep{Crooks2007,Feng2009}, which requires significant amount
of small steps. In this Letter, we define a finite-time thermodynamic
length $\mathcal{L}(\tau),$ which retains the thermodynamic length
at the long-time limit. The thermodynamic length $\mathcal{L}$ can
be measured through extrapolating few data points of $\mathcal{L}(\tau)$
with finite duration $\tau$. The measurement procedure is illustrated
via examples of the quantum harmonic oscillator with tuning frequency
and the classical ideal gas with changing volume.

\textit{Finite-time thermodynamic length.}---We consider an open
quantum system with the control parameter $\lambda(t)$ tuned from
the initial value $\lambda(0)=\lambda_{0}$ to the final value $\lambda(\tau)=\lambda_{\tau}$
in a finite-time process with duration $\tau$. The system state is
described by the density matrix $\rho(t)$ $(0<t<\tau)$, which evolves
under the time-dependent Hamiltonian $H(t)=H[\lambda(t)]$ via the
time-dependent Markovian master equation \citep{Albash2012,Yamaguchi2017,Dann2018}
\begin{equation}
\dot{\rho}=\mathscr{L}_{\lambda(t)}\rho,\label{eq:master_equation}
\end{equation}
where $\mathscr{L}_{\lambda(t)}$ is the quantum Liouvillian super-operator.
In quantum thermodynamics, the rate of the performed work, namely
the power, is $\dot{W}=\mathrm{Tr}(\rho\dot{H})$ \citep{Alicki1979}.
In a quasi-static isothermal process with infinite duration, the system
evolves along the trajectory of the equilibrium state $\rho_{\mathrm{eq}}(t)=\exp[-\beta_{\mathrm{b}}H(t)]/\mathrm{Tr\{}\exp[-\beta_{\mathrm{b}}H(t)]\}$
with the inverse temperature $\beta_{\mathrm{b}}=1/k_{\mathrm{B}}T_{\mathrm{b}}$
of the bath, and the performed work of the whole process is $W_{(0)}=\int_{\lambda_{0}}^{\lambda_{\tau}}\mathrm{\mathrm{Tr}}(\rho_{\mathrm{eq}}\partial H/\partial\lambda)d\lambda$.
In a finite-time isothermal process, the excess work $W_{\mathrm{ex}}(\tau)=\int_{0}^{\tau}P_{\mathrm{ex}}(t)dt$
is utilized to evaluate the dissipation with the excess power \citep{Sivak2012}
$P_{\mathrm{ex}}(t)=\dot{W}(t)-\dot{W}_{(0)}(t),$ where $\dot{W}_{(0)}(t)=\mathrm{\mathrm{Tr}}[\rho_{\mathrm{eq}}\dot{H}]$
is the quasi-static part of the work rate. One important progress
\citep{Sivak2012,Zulkowski2012} for the finite-time thermodynamics
is the discovery of the geometric bound of the excess work with the
thermodynamic length as $W_{\mathrm{ex}}(\tau)\geq\mathcal{L}^{2}/\tau$,
and the equality is saturated by the optimal protocol with the constant
excess power. The direct measurement of thermodynamic length $\mathcal{L}$
requires the infinite slow isothermal processes \citep{Crooks2007,Feng2009}.

In quantum thermodynamics, the thermodynamic length $\mathcal{L}$
is explicitly \citep{Scandi2018}
\begin{equation}
\mathcal{L}=\int_{\lambda_{0}}^{\lambda_{\tau}}\sqrt{\mathrm{\mathrm{Tr}}[\frac{\partial H}{\partial\lambda}\mathscr{L}_{\lambda}^{-1}(\frac{\partial\rho_{\mathrm{eq}}}{\partial\lambda})]}d\lambda,\label{eq:quantum_thermodynamic_length}
\end{equation}
which contains the Drazin inverse $\mathscr{L}_{\lambda}^{-1}$ of
the super-operator \citep{Mandal2016,ondrazininverse}. For the diagonalizable
super-operator, the Drazin inverse is explicitly obtained as follows.
The eigendecomposition is $\mathscr{L}_{\lambda}=\sum_{\Gamma}\Gamma P_{\Gamma}$
with the eigenvalues $\Gamma$ and the projections $P_{\Gamma}$ in
the super-space of the density matrices. The null eigenvector is the
instantaneous equilibrium state satisfied $\mathscr{L}_{\lambda}[\rho_{\mathrm{eq}}(\lambda)]=0$.
The Drazin inverse is then written as $\mathscr{L}_{\lambda}^{-1}=\sum_{\Gamma\ne0}\Gamma^{-1}P_{\Gamma}$,
where the inverses of the non-zero eigenvalues $\Gamma^{-1}$ determines
different dissipation timescales \citep{Mandal2016,ondrazininverse}.
Detailed discussions about the Drazin inverse are given with an example
of the two-level system in the Supplementary Materials \citep{supportingmaterialsec3}.

For the measurement, we define a finite-time thermodynamic length
as 
\begin{equation}
\mathcal{L}(\tau)=\int_{0}^{\tau}\sqrt{P_{\mathrm{ex}}(t)}dt.\label{eq:finite_time_thermodynamic_length}
\end{equation}
The two following properties of the finite-time thermodynamic length
allows the measurement of the thermodynamics length with the extrapolation
of finite-time measurements.

(i) \textbf{The convergence $\underset{\tau\rightarrow\infty}{\lim}\mathcal{L}(\tau)=\mathcal{L}$}.
In a slow isothermal process, the state of the system evolves near
the equilibrium state, and the solution to Eq. (\ref{eq:master_equation})
is expanded in the series \citep{Cavina2017} as

\begin{equation}
\rho(t)=\sum_{n=0}^{\infty}\left(\mathscr{L}_{\lambda(t)}^{-1}\frac{d}{dt}\right)^{n}\rho_{\mathrm{eq}}(t).\label{eq:series-rho(t)}
\end{equation}
For a given tuning protocol $\lambda(t)=\tilde{\lambda}(t/\tau)$,
the series expansion of the excess power is obtained as

\begin{equation}
P_{\mathrm{ex}}(t)=\mathrm{\mathrm{Tr}}\{\frac{\partial\tilde{H}(s)}{\partial s}\sum_{n=1}^{\infty}\tau^{1-n}(\mathscr{L}_{\tilde{\lambda}(s)}^{-1}\frac{\partial}{\partial s})^{n}[\tilde{\rho}_{\mathrm{eq}}(s)]\},\label{eq:excesspower}
\end{equation}
where $s=t/\tau$ is the rescaled dimensionless time. Equation (\ref{eq:excesspower})
is invalid at the beginning of the isothermal process. Its validity
requires a relaxation with the time larger than the typical dissipation
timescale. For the slow process, the lowest-order term with $n=1$
dominates the summation, and the finite-time thermodynamic length
approaches the thermodynamic length at the long-time limit.

(ii) \textbf{The protocol independence of the limit $\underset{\tau\rightarrow\infty}{\lim}\mathcal{L}(\tau)$}.
For the system with only one control parameter, the limit of the finite-time
thermodynamic length $\underset{\tau\rightarrow\infty}{\lim}\mathcal{L}(\tau)$
is independent of the protocol, which implies the thermodynamic length
can be measured without necessarily using the optimal protocol. For
multiple control parameters, the limit $\underset{\tau\rightarrow\infty}{\lim}\mathcal{L}(\tau)$
indeed relies on the path of the protocol. The minimal thermodynamic
length is only reached by the geodesic path, and the optimized protocol
is to tune the control parameter with the constant velocity of the
thermodynamic length \citep{Scandi2018}.

\begin{figure}
\includegraphics[width=7cm]{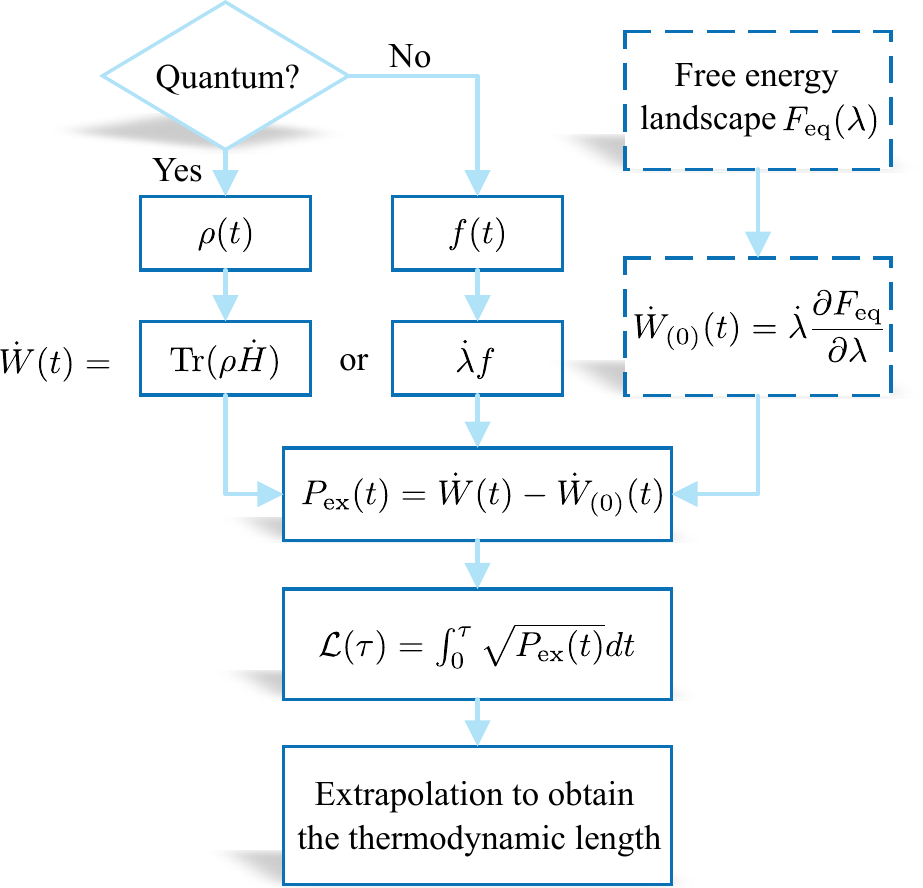}

\caption{(Color online) Flowchart for measuring the thermodynamic length $\mathcal{L}$
with finite-time extrapolation. The excess power is evaluated with
the performed work rate $\dot{W}(t)$ and the quasi-static work rate
$\dot{W}_{(0)}(t)$ at each moment of the finite-time process.}

\label{fig:strategy}
\end{figure}

\textit{Measurement strategy.}---The thermodynamic length is the
long-time limit of the finite-time thermodynamic length. In the following,
we propose an extrapolation method to measure the thermodynamic length
for the system with a single control parameter. The flowchart of the
method is shown in Fig. \ref{fig:strategy}. The measurement of the
finite-time thermodynamic length $\mathcal{L}(\tau)$ requires the
excess power $P_{\mathrm{ex}}(t)$ of the whole process. The power
$\dot{W}(t)\;(0<t<\tau)$ in a finite-time process is measured through
the conjugate force $f(t)=\mathrm{\mathrm{Tr}}[\rho\partial_{\lambda}H]$
for a classical system via $\dot{W}(t)=\dot{\lambda}(t)f(t)$ or the
tomography of the density matrix $\rho(t)$ for a quantum system via
$\dot{W}(t)=\mathrm{Tr}(\rho\dot{H})$. With the given protocol $\tilde{\lambda}(s)$
of the control parameter, the quasi-static work rate $\dot{W}_{(0)}(t)=\mathrm{Tr}(\rho_{\mathrm{eq}}\dot{H})=\dot{\lambda}\partial_{\lambda}F_{\mathrm{eq}}(\lambda)$
is obtained with the landscape of the equilibrium free energy $F_{\mathrm{eq}}(\lambda)$,
which can be typically obtained from finite-time processes via Jarzynski's
equality \citep{Jarzynski1997}.

Assuming the length $\mathcal{L}(\tau)$ as a smooth function of the
duration $\tau$, we expand $\mathcal{L}(\tau)$ with the Laurent
series as

\begin{equation}
\mathcal{L}(\tau)=\mathcal{L}+\sum_{j=1}^{\infty}\frac{a_{j}}{\tau^{j}}.
\end{equation}
The duration $\tau$ needs to be chosen notably larger than the dissipation
timescale to ensure the validity of Eq. (\ref{eq:excesspower}). By
measuring the finite-time thermodynamic length under given duration,
we extrapolate the function as $\mathcal{L}(\tau)=\mathcal{L}+\sum_{j=1}^{N}a_{j}\tau^{-j}$
with the cutoff $N$. The thermodynamic length $\mathcal{L}$ is estimated
with the extrapolation $\tau\rightarrow\infty$. We apply the current
strategy to measure the thermodynamic length with two examples, i.e.
the quantum harmonic oscillator and the classical ideal gas system.

\textit{Quantum harmonic oscillator with tuned frequency.}---We consider
the quantum Brownian motion with the Hamiltonian $H(t)=\hat{p}^{2}/(2m)+m\omega(t)^{2}\hat{x}^{2}/2$
in a tuned harmonic potential with the frequency $\omega(t)$ as the
control parameter in the finite-time isothermal process. At the high
temperature limit, the evolution of the reduced system is described
by the Caldeira-Leggett master equation \citep{Caldeira1983,Breuer2007},
i.e, $\partial_{t}\rho=\mathscr{L}_{\omega(t)}\rho$. The quantum
Liouvillian super-operator is explicitly

\begin{equation}
\mathscr{L}_{\omega(t)}\rho=-i[H(t),\rho]-i\kappa[\hat{x},\{\hat{p},\rho\}]-\frac{2m\kappa}{\beta_{\mathrm{b}}}[\hat{x},[\hat{x},\rho]],\label{eq:brownianmotionequation}
\end{equation}
where the frequency-independent damping rate $\kappa$ is induced
by the Ohmic spectral of the heat bath \citep{Caldeira1983,Breuer2007}.

With the infinite dimension of the Hilbert space for a harmonic oscillator,
it is difficult to solve the evolution of the density matrix by calculating
the Drazin inverse of the super-operator directly. An alternative
way is to solve the finite-time dynamics via a closed Lie algebra
\citep{Rezek2006,Lee2020} with the thermodynamic variables, the Hamiltonian
$H(t)$, the Lagrange $L(t)=\hat{p}^{2}/(2m)-m\omega(t)^{2}\hat{x}^{2}/2$
and the correlation function $D(t)=\omega(t)(\hat{x}\hat{p}+\hat{p}\hat{x})/2$.
The closed differential equations of the expectations of the thermodynamic
variables $\left\langle H(t)\right\rangle =\mathrm{Tr}[\rho(t)H(t)]$,
$\left\langle L(t)\right\rangle =\mathrm{Tr}[\rho(t)L(t)]$ and $\left\langle D(t)\right\rangle =\mathrm{Tr}[\rho(t)D(t)]$
are obtained from Eq. (\ref{eq:brownianmotionequation}) as
\begin{widetext}
\begin{equation}
\frac{d}{dt}\left(\begin{array}{c}
\left\langle H\right\rangle \\
\left\langle L\right\rangle \\
\left\langle D\right\rangle 
\end{array}\right)=\left(\begin{array}{ccc}
-2\kappa+\frac{\dot{\omega}}{\omega} & -2\kappa-\frac{\dot{\omega}}{\omega} & 0\\
-2\kappa-\frac{\dot{\omega}}{\omega} & -2\kappa+\frac{\dot{\omega}}{\omega} & -2\omega\\
0 & 2\omega & -2\kappa+\frac{\dot{\omega}}{\omega}
\end{array}\right)\left(\begin{array}{c}
\left\langle H\right\rangle \\
\left\langle L\right\rangle \\
\left\langle D\right\rangle 
\end{array}\right)+2\kappa k_{B}T_{\mathrm{b}}\left(\begin{array}{c}
1\\
1\\
0
\end{array}\right).\label{eq:differential}
\end{equation}
The derivation to Eqs. (\ref{eq:differential}) is presented in the
Supplementary Materials \citep{supportingmaterialsec3}.
\end{widetext}

The performed work rate of the tuned harmonic oscillator is $\dot{W}=\dot{\omega}/\omega\left(\left\langle H\right\rangle -\left\langle L\right\rangle \right)$.
In a quasi-static isothermal process, the system evolves along the
equilibrium states with the average internal energy $\left\langle H\right\rangle =k_{B}T_{\mathrm{b}}$
and the average Lagrange $\left\langle L\right\rangle =0$, and the
quasi-static work rate is $\dot{W}_{(0)}(t)=k_{B}T_{\mathrm{b}}\dot{\omega}/\omega$.
In a finite-time process, the finite-time thermodynamic length $\mathcal{L}(\tau)$
is explicitly
\begin{equation}
\mathcal{L}(\tau)=\int_{0}^{\tau}\left[\dot{\omega}/\omega\left(\left\langle H\right\rangle -\left\langle L\right\rangle -k_{B}T_{\mathrm{b}}\right)\right]^{1/2}dt.
\end{equation}

The thermodynamic length of the tuned harmonic oscillator is obtained
by Eq. (\ref{eq:quantum_thermodynamic_length}) as

\begin{align}
\mathcal{L} & =\sqrt{\frac{k_{B}T_{\mathrm{b}}}{2\kappa}}\left|\left[\text{sinh}^{-1}(\frac{\omega}{2\kappa})-\sqrt{1+\frac{4\kappa^{2}}{\omega^{2}}}\right]\Big|_{\omega=\omega_{0}}^{\omega_{\tau}}\right|,\label{eq:harmoniclength}
\end{align}
where $\omega_{0}$ and $\omega_{\tau}$ are the initial and final
frequencies of the harmonic potential. The detailed derivation of
the thermodynamic length is given in the Supplementary Materials \citep{supportingmaterialsec3}.
The optimal protocol $\omega_{\mathrm{op}}(t)=\tilde{\omega}_{\mathrm{op}}(t/\tau)$
satisfies

\begin{equation}
\left(\frac{1}{2\kappa}+\frac{2\kappa}{\tilde{\omega}_{\mathrm{op}}^{2}}\right)^{1/2}\frac{1}{\tilde{\omega}_{\mathrm{op}}}\frac{d\tilde{\omega}_{\mathrm{op}}(s)}{ds}=\mathrm{const}.\label{eq:optimalprotocol}
\end{equation}
For small damping rate $\kappa\ll\omega$, the thermodynamic length
approximates $\mathcal{L}\approx\sqrt{k_{B}T_{\mathrm{b}}/(2\kappa)}\left|\ln\omega_{\tau}/\omega_{0}\right|$,
and the optimal protocol is the exponential protocol $\tilde{\omega}_{\mathrm{op}}(s)=\omega_{0}\exp[s\ln(\omega_{\tau}/\omega_{0})]$.
For the large damping rate $\omega\ll\kappa$, the thermodynamic length
approximates $\mathcal{L}\approx\sqrt{2\kappa k_{B}T_{\mathrm{b}}}\left|(\omega_{\tau}-\omega_{0})/(\omega_{0}\omega_{\tau})\right|$,
and the optimal protocol is the inverse protocol $\tilde{\omega}_{\mathrm{op}}(s)=\omega_{0}\omega_{\tau}/[(\omega_{0}-\omega_{\tau})s+\omega_{\tau}]$.

\begin{figure}
\includegraphics[width=7cm]{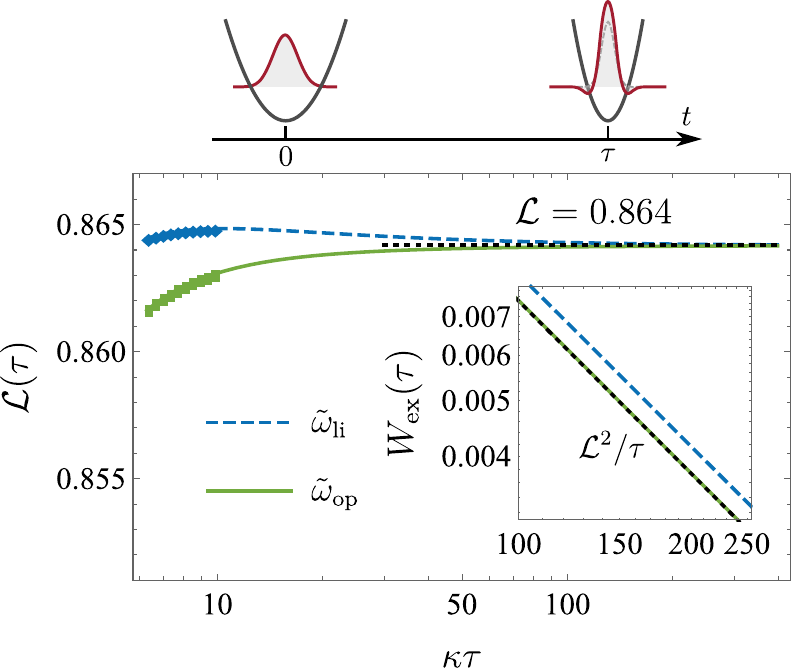}

\caption{(Color online) Finite-time extrapolation to measure the thermodynamic
length for the quantum Brownian motion in a tuned harmonic potential.
The upper panel shows the frequency increases from $\omega_{0}=1$
to $\omega_{\tau}=2$. The temperature and the damping rate are set
as $k_{B}T_{\mathrm{b}}=1$ and $\kappa=1$. In the lower panel, the
black dotted line presents the thermodynamic length $\mathcal{L}=0.864$,
which is approached by the extrapolated functions with $N=3$ for
both the linear (blue dashed curve) and the optimal (green solid curve)
protocols. The subset shows the excess work in a slow process is bounded
as $W_{\mathrm{ex}}(\tau)\protect\geq\mathcal{L}^{2}/\tau$ (black
dotted line), which is saturated by the optimal protocol at the long-time
limit. \label{fig:length=000026excess work}}
\end{figure}

We exemplify the finite-time extrapolation method to obtain the thermodynamic
length of the tuned harmonic oscillator through numerically solving
the relaxation dynamics with the linear protocol $\tilde{\omega}_{\mathrm{li}}(s)=\omega_{0}+(\omega_{\tau}-\omega_{0})s$
and the optimal protocol $\tilde{\omega}_{\mathrm{op}}(s)$. In the
numerical calculation, the frequency is tuned from $\omega_{0}=1$
to $\omega_{\tau}=2$ with the temperature $k_{B}T_{\mathrm{b}}=1$
and the damping rate $\kappa=1$. In Fig. \ref{fig:length=000026excess work},
the extrapolated functions (curves) with $N=3$ are obtained from
10 sets of data (markers) with the duration $\tau$ ranging from $6.4$
to $9.9$ for the linear (blue dashed line) and the optimal (the green
solid line) protocols as $\mathcal{L}_{\mathrm{li}}(\tau)=0.864+0.0128/\tau-0.0464/\tau^{2}-0.145/\tau^{3}$
and $\mathcal{L}_{\mathrm{op}}(\tau)=0.864-0.00471/\tau-0.0448/\tau^{2}-0.168/\tau^{3}$.
The two extrapolated functions both give the consistent thermodynamic
length identical to the theoretical result $\mathcal{L}=0.864$ by
Eq. (\ref{eq:harmoniclength}), as illustrated with the dotted black
line. Therefore the current extrapolation method enables the measurement
of the thermodynamic length in relatively short time without the need
to find the optimal protocol.

The subset shows the $\tau^{-1}$ scaling of the excess work. The
black dotted line shows the bound $\mathcal{L}^{2}/\tau$, and the
blue dashed (green solid) line shows the extra work with the linear
(optimal) control scheme. The excess work of different protocols is
indeed bounded by the thermodynamic length, and the bound is saturated
by the optimal protocol.

\textit{Compression of classical ideal gas.}--- The extrapolation
method is applicable in classical systems with the strategy shown
in Fig. \ref{fig:strategy}. We consider the finite-time compression
of the classical ideal gas in a box, which is in contact with a heat
bath at the temperature $T_{\mathrm{b}}$. By compressing the piston,
the volume of the box changes with the performed work rate $\dot{W}=-p\dot{V}$.
With the state equation $pV=nRT$, the temperature $T$ of the classical
ideal gas satisfies 
\begin{equation}
\frac{dT}{dt}=-\frac{nRT}{C_{V}}\frac{1}{V}\frac{dV}{dt}-\gamma(T-T_{\mathrm{b}}),\label{eq:equationofT}
\end{equation}
where $\gamma$ is the cooling rate in the Newton's law of cooling,
assumed as a constant in the following discussion, and $C_{V}$ is
the heat capacity at the constant volume, e.g. $C_{V}=3nR/2$ for
the single-atom ideal gas. In a finite-time isothermal process, the
excess power is obtained as $P_{\mathrm{ex}}(t)=-nR(T-T_{\mathrm{b}})\dot{V}/V$.
The thermodynamic length $\mathcal{L}$ is theoretically obtained
as

\begin{equation}
\mathcal{L}=\sqrt{\frac{(nR)^{2}T_{\mathrm{b}}}{\gamma C_{V}}}\left|\ln\frac{V_{\tau}}{V_{0}}\right|.\label{eq:gaslength}
\end{equation}
For a long-time compression, the optimal protocol with the constant
excess power is obtained as the exponential protocol $\tilde{V}_{\mathrm{op}}(s)=V_{0}(V_{\tau}/V_{0})^{s}$,
which is consistent with the result obtained by the stochastic thermodynamics
\citep{Gong2016}.

\begin{figure}
\includegraphics[width=7cm]{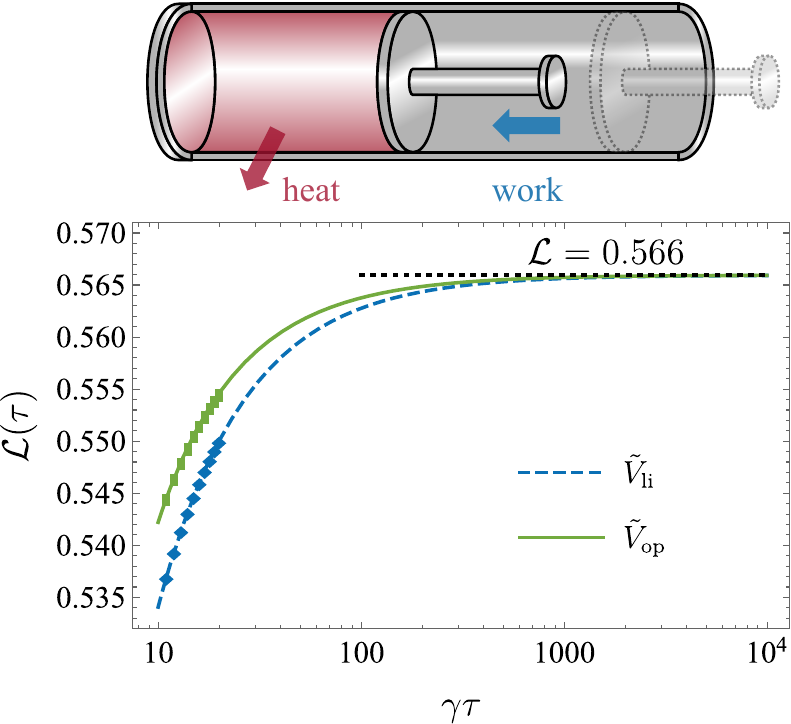}

\caption{(Color online) Finite-time extrapolation to measure the thermodynamic
length for the compression of the classical ideal gas in a box. The
upper panel shows the compression of the box. In the lower panel,
we use the linear protocol (blue dashed curve) and the optimal protocol
(green solid curve) to compress the volume of the box to one half.
With the current chosen parameters, the length extrapolated with the
two protocols match the exact thermodynamic length $\mathcal{L}=0.566$
(black dotted line) by Eq. (\ref{eq:gaslength}). \label{fig:extrapolationofidealgas}}
\end{figure}

In Fig. \ref{fig:extrapolationofidealgas}, the finite-time thermodynamic
lengths (markers) for given protocols are numerically obtained for
the isothermal processes with the duration $\tau$ ranging from $11.0$
to $20.0$, where the parameters are set as $nR=T_{\mathrm{b}}=\gamma=1$
and $C_{V}=1.5$, and the volume is tuned from $V_{0}=1$ to $V_{\tau}=0.5$.
Two protocols are considered to tune the volume linearly $\tilde{V}_{\mathrm{li}}(s)=1-s/2$
(blue dashed curve) and exponentially $\tilde{V}_{\mathrm{op}}(s)=2^{-s}$
(green solid curve). Setting the cutoff $N=3$, the extrapolated functions
of the two protocols are $\mathcal{L}_{\mathrm{li}}(\tau)=0.566-0.318/\tau+0.010/\tau^{2}-0.146/\tau^{3}$
and $\mathcal{L}_{\mathrm{op}}(\tau)=0.566-0.216/\tau-0.199/\tau^{2}-0.111/\tau^{3}$.
Both extrapolations with different protocols lead to the same value,
matching the theoretical result $\mathcal{L}=0.566$ (black dotted
line) given by Eq. (\ref{eq:gaslength}).

Using examples of both the quantum harmonic oscillator and the classical
ideal gas, we demonstrate the proposed extrapolation method for measuring
the thermodynamic length for typical systems with single control parameter.
The quantum harmonic oscillator system is potentially realized with
the trapped ion \citep{An2014Experimental,Rossnagel2016}, and the
classical ideal gas system will be tested with our recently designed
apparatus \citep{Ma2019}.

\textit{Conclusion.}---The thermodynamic length is crucial in optimization
of finite-time isothermal processes, yet remains challenging to be
practically measured in experiments. We have overcome the difficulty
and proposed to measure the thermodynamic length via the finite-time
extrapolation method, which allows to determine the lower bound of
the excess work without necessarily using the optimal protocol. The
protocol-independent property of the current measurement strategy
benefits to practically measure the thermodynamic length in experiments.
Based on our recent experimental setup \citep{Ma2019}, it is feasible
to measure the thermodynamic length for the finite-time compression
of the classical ideal gas in the experiments, which we leave as the
goal for our future experimental research.
\begin{acknowledgments}
Jin-Fu Chen thank Ruo-Xun Zhai for the helpful discussions about the
dry-air experiment. This work is supported by the National Natural
Science Foundation of China (NSFC) (Grants No. 11534002, No. 11875049,
No. U1930402, and No. U1930403), and the National Basic Research Program
of China (Grants No. 2016YFA0301201 and No. 2014CB921403).
\end{acknowledgments}

\bibliographystyle{apsrev4-1}
\bibliography{mainref}

\end{document}


\title{Supplementary Materials: Extrapolating the thermodynamic length with
finite-time measurements}
\author{Jin-Fu Chen}
\address{Beijing Computational Science Research Center, Beijing 100193, China}
\address{Graduate School of China Academy of Engineering Physics, No. 10 Xibeiwang
East Road, Haidian District, Beijing, 100193, China}
\author{Chang-Pu Sun}
\email{cpsun@gscaep.ac.cn}

\address{Beijing Computational Science Research Center, Beijing 100193, China}
\address{Graduate School of China Academy of Engineering Physics, No. 10 Xibeiwang
East Road, Haidian District, Beijing, 100193, China}
\author{Hui Dong}
\email{hdong@gscaep.ac.cn}

\address{Graduate School of China Academy of Engineering Physics, No. 10 Xibeiwang
East Road, Haidian District, Beijing, 100193, China}
\date{\today}

\maketitle
The document is devoted to providing detailed derivations and supporting
discussions to the main content. In Sec. \ref{sec:Quantum-thermodynamic-length},
we show the convergence of the finite-time thermodynamic length to
the thermodynamic length with the increasing process time $\tau$.
In Sec. \ref{sec:two-level-system-with}, we show the example of the
tuned two-level system, and optimize the protocol with the Drazin
inverse of the Lindblad operator. In Sec. \ref{sec:Case-study:-quantum},
we show detailed discussions for the tuned harmonic oscillator, especially
the derivation of Eq. (8) in the main content. In Sec. \ref{sec:Case-study:-Compression},
we consider the thermodynamic length for the pedagogical model of
the classical ideal gas system, which was used to validate the $\tau^{-1}$
scaling of the excess work in our recent experiment \citep{Ma2019}.

\section{Convergence of the finite-time thermodynamic length\label{sec:Quantum-thermodynamic-length}}

We demonstrate the convergence of the finite-time thermodynamic length
\textbf{$\mathcal{L}=\underset{\tau\rightarrow\infty}{\lim}\mathcal{L}(\tau)$}.
Plugging into the series expansion of the excess power by Eq. (5)
in the main content, the finite-time thermodynamic length is rewritten
as 
\begin{align}
\mathcal{L}(\tau) & =\int_{0}^{1}\sqrt{\sum_{n=1}^{\infty}\mathrm{\mathrm{Tr}}\{\frac{\partial\tilde{H}(s)}{\partial s}\tau^{1-n}(\mathscr{L}_{\tilde{\lambda}(s)}^{-1}\frac{\partial}{\partial s})^{n}[\tilde{\rho}_{\mathrm{eq}}(s)]\}}ds.
\end{align}
With the increasing control time $\tau$, the term with $n=1$ is
independent of $\tau$ and dominates the summation, and leads to the
thermodynamic length 
\begin{align}
\mathcal{L} & =\int_{0}^{1}\sqrt{\mathrm{\mathrm{Tr}}\{\frac{\partial\tilde{H}(s)}{\partial s}(\mathscr{L}_{\tilde{\lambda}(s)}^{-1}\frac{\partial}{\partial s})[\tilde{\rho}_{\mathrm{eq}}(s)]\}}ds,\\
 & =\int_{0}^{1}\sqrt{[\tilde{\lambda}^{\prime}(s)]^{2}\mathrm{\mathrm{Tr}}\left[\left(\frac{\partial H}{\partial\lambda}\right)\left(\mathscr{L}_{\tilde{\lambda}(s)}^{-1}\frac{\partial\tilde{\rho}_{\mathrm{eq}}}{\partial\lambda}\right)\right]}ds,\\
 & =\int_{\lambda_{0}}^{\lambda_{1}}\sqrt{\mathrm{\mathrm{Tr}}\left(\frac{\partial H}{\partial\lambda}\right)\left(\mathscr{L}_{\lambda}^{-1}\frac{\partial\rho_{\mathrm{eq}}}{\partial\lambda}\right)}\left|d\lambda\right|.
\end{align}
The last integral shows $\mathcal{L}$ is independent of the protocol
$\tilde{\lambda}(s)$ for tuning a single control parameter.

\section{two-level system with tuned energy spacing \label{sec:two-level-system-with}}

In this section, we show the example of two-level system with tuned
energy spacing. The Drazin Inverse of the Lindblad operator is obtained
directly for the optimization of the protocol. The system Hamiltonian
of the two-level system reads

\begin{equation}
H(t)=\frac{E(t)}{2}\left(\left|\mathrm{e}\right\rangle \left\langle \mathrm{e}\right|-\left|\mathrm{g}\right\rangle \left\langle \mathrm{g}\right|\right),
\end{equation}
where the energy spacing $E(t)$ is the control parameter. The state
of the two-level system is represented with the density matrix

\begin{equation}
\rho(t)=\left(\begin{array}{cc}
\rho_{\mathrm{ee}} & \rho_{\mathrm{eg}}\\
\rho_{\mathrm{ge}} & \rho_{\mathrm{gg}}
\end{array}\right).
\end{equation}

With the coupling to the heat bath, the evolution is governed by the
time-dependent master equation 
\begin{equation}
\dot{\rho}=\mathscr{L}_{E(t)}(\rho),\label{eq:6}
\end{equation}
where $\mathscr{L}_{E(t)}$ is in the Lindblad form

\begin{align}
\mathscr{L}_{E(t)}(\rho) & =-i[H(t),\rho]+\gamma_{\uparrow}(t)\mathcal{D}(\sigma_{+})[\rho]+\gamma_{\downarrow}(t)\mathcal{D}(\sigma_{-})[\rho],\label{eq:Lindbladoperatortwolevel-1}
\end{align}
with the transition operators $\sigma_{+}=\left|\mathrm{e}\right\rangle \left\langle \mathrm{g}\right|,\sigma_{-}=\left|\mathrm{g}\right\rangle \left\langle \mathrm{e}\right|$
and the dissipation super-operator

\begin{equation}
\mathcal{D}(\sigma)[\rho]=\sigma\rho\sigma^{\dagger}-\frac{1}{2}\sigma^{\dagger}\sigma\rho-\frac{1}{2}\rho\sigma^{\dagger}\sigma.
\end{equation}
The time-dependent transition rates are $\gamma_{\uparrow}(t)=\gamma(t)N(t)$
and $\gamma_{\downarrow}(t)=\gamma(t)[N(t)+1]$ with the average phonon
number

\begin{equation}
N(t)=\frac{1}{e^{\beta_{\mathrm{b}}E(t)}-1},
\end{equation}
The spontaneous emission rate $\gamma(t)$ relies on the bath spectral
as

\begin{equation}
\gamma(t)=\gamma_{0}E(t)^{\alpha}.
\end{equation}
Equation (\ref{eq:Lindbladoperatortwolevel-1}) extends the well-known
Lindblad master equation \citep{Breuer2007} for the tuned two-level
system via the time-dependent energy spacing $E(t)$. This approximation
is suitable for a long-time isothermal process with slowly tuned control
parameters.

Rewriting the density matrix into a vector

\begin{equation}
\rho=\left(\begin{array}{cccc}
\rho_{\mathrm{ee}} & \rho_{\mathrm{gg}} & \rho_{\mathrm{eg}} & \rho_{\mathrm{ge}}\end{array}\right)^{\mathrm{T}},
\end{equation}
the Lindblad operator is presented in the matrix form as

\begin{equation}
\mathscr{L}_{E(t)}=\left(\begin{array}{cccc}
-\gamma(N+1) & \gamma N & 0 & 0\\
\gamma(N+1) & -\gamma N & 0 & 0\\
0 & 0 & -\gamma(N+\frac{1}{2})-iE & 0\\
0 & 0 & 0 & -\gamma(N+\frac{1}{2})+iE
\end{array}\right),
\end{equation}
with the eigenvalues $\Gamma=0,-\gamma(N+1/2)\pm iE$, and $\gamma(2N+1)$.
The Drazin inverse of the Lindblad operator is obtained as

\begin{align}
\mathscr{L}_{E(t)}^{-1} & =\left(\begin{array}{cccc}
-\frac{(N+1)}{\gamma(2N+1)^{2}} & \frac{N}{\gamma(2N+1)^{2}} & 0 & 0\\
\frac{(N+1)}{\gamma(2N+1)^{2}} & -\frac{N}{\gamma(2N+1)^{2}} & 0 & 0\\
0 & 0 & \frac{1}{-\gamma(N+\frac{1}{2})-iE} & 0\\
0 & 0 & 0 & \frac{1}{-\gamma(N+\frac{1}{2})+iE}
\end{array}\right).
\end{align}

At the time $t$, the instantaneous equilibrium state is 
\begin{align}
\rho_{\mathrm{eq}}(t) & =\left(\begin{array}{cccc}
\frac{N(t)}{2N(t)+1} & \frac{N(t)+1}{2N(t)+1} & 0 & 0\end{array}\right)^{\mathrm{T}}.
\end{align}
The off-diagonal terms of the density matrix remain zero $\rho_{\mathrm{eg}}(t)=\rho_{\mathrm{ge}}^{*}(t)=0$
during the whole isothermal process. The excess power to the first
order is

\begin{align}
P_{\mathrm{ex}}^{[1]}(t) & =\mathrm{Tr}[\frac{dH(t)}{dt}\mathscr{L}_{E(t)}^{-1}(\frac{d\rho_{\mathrm{eq}}(t)}{dt})]\\
 & =\frac{\beta_{b}\dot{E}^{2}}{4\gamma}\frac{\tanh\left(\frac{1}{2}\beta_{b}E\right)}{\cosh^{2}\left(\frac{1}{2}\beta_{b}E\right)}.
\end{align}
The thermodynamic length follows as 
\begin{equation}
\mathcal{L}=\int_{E_{0}}^{E_{\tau}}\sqrt{\frac{\beta_{\mathrm{b}}}{4\gamma_{0}E^{\alpha}}\frac{\tanh\left(\frac{1}{2}\beta_{\mathrm{b}}E(t)\right)}{\cosh^{2}\left(\frac{1}{2}\beta_{\mathrm{b}}E(t)\right)}}dE.\label{eq:Lengthfixeddirection}
\end{equation}
 At the long-time limit, the optimal protocol $\tilde{E}(s)$ is obtained
with the constant excess power $P_{\mathrm{ex}}^{[1]}(t)=\mathrm{const}$,
namely
\begin{equation}
\frac{\beta_{\mathrm{b}}}{4\gamma_{0}\tilde{E}(s)^{\alpha}}\frac{\tanh\left(\frac{1}{2}\beta_{\mathrm{b}}\tilde{E}(s)\right)}{\cosh^{2}\left(\frac{1}{2}\beta_{\mathrm{b}}\tilde{E}(s)\right)}\left(\frac{d\tilde{E}(s)}{ds}\right)^{2}=\mathrm{const}.\label{eq:optimalprotocol}
\end{equation}

\begin{figure}[h]
\includegraphics[width=7cm]{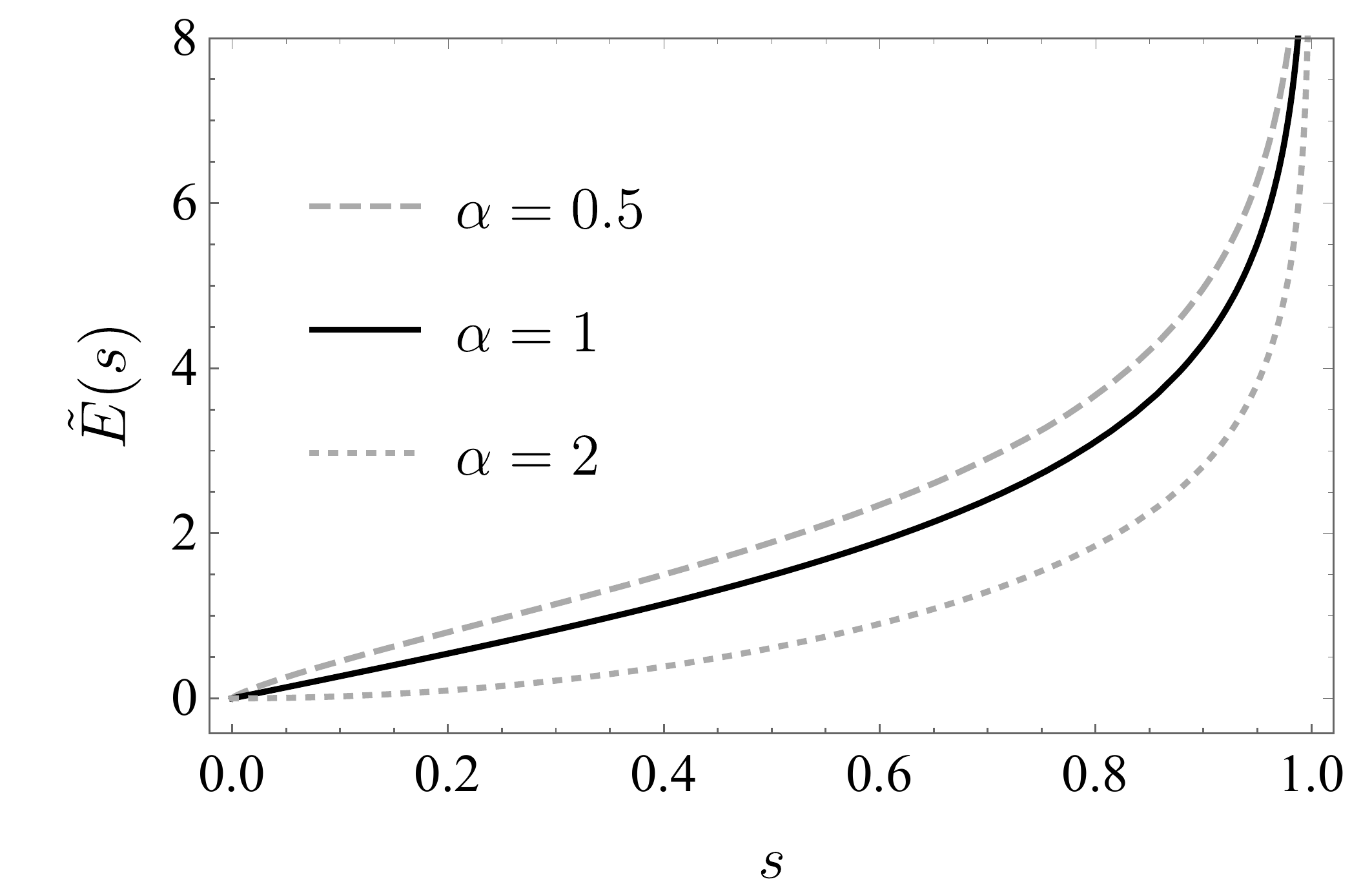}

\caption{The optimal protocols of tuning the energy spacing $\tilde{E}(s)$
of the two-level system with different bath spectral $\alpha=0.5,\,1$
and $2$. The optimal protocols are obtained by numerically solving
Eq. (\ref{eq:optimalprotocol}) with the parameters $\beta_{\mathrm{b}}=1$
and $\gamma_{0}=1$. \label{fig:The-optimal-protocol}}
\end{figure}

Figure \ref{fig:The-optimal-protocol} shows the tuning of the energy
spacing $\tilde{E}(s)$ with $s$ ranging from $0$ to $1$. We solve
the optimal protocols for different bath spectral, the sub-Ohmic $\alpha=0.5$
(the dashed curve), the Ohmic $\alpha=1$ (the dotted curve), and
super-Ohmic $\alpha=2$ (the dash-dotted curve) with the inverse temperature
$\beta_{\mathrm{b}}=1$ and the dissipation strength $\gamma_{0}=1$.

\begin{figure}[h]
\includegraphics[width=7cm]{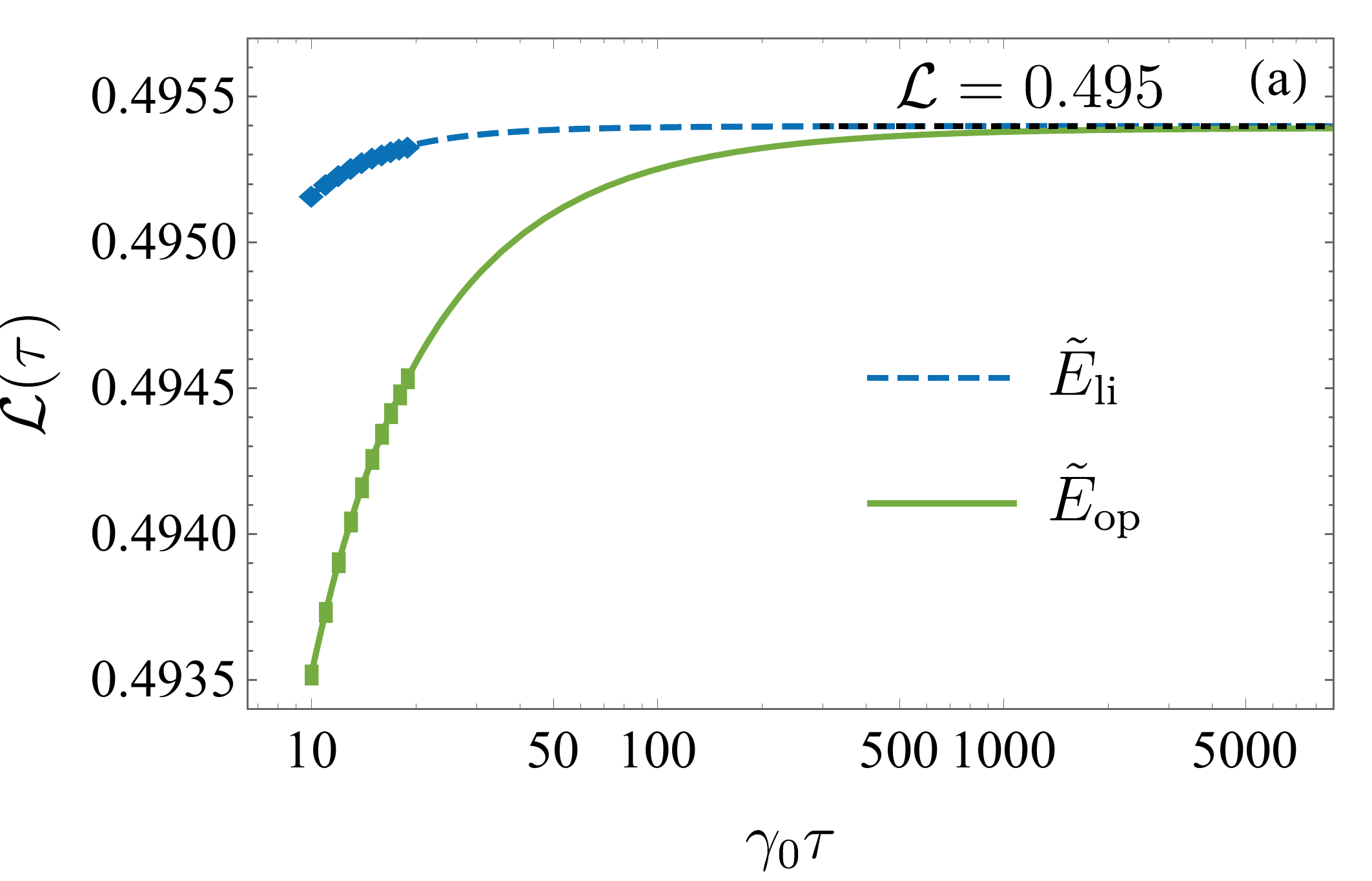}\includegraphics[width=7cm]{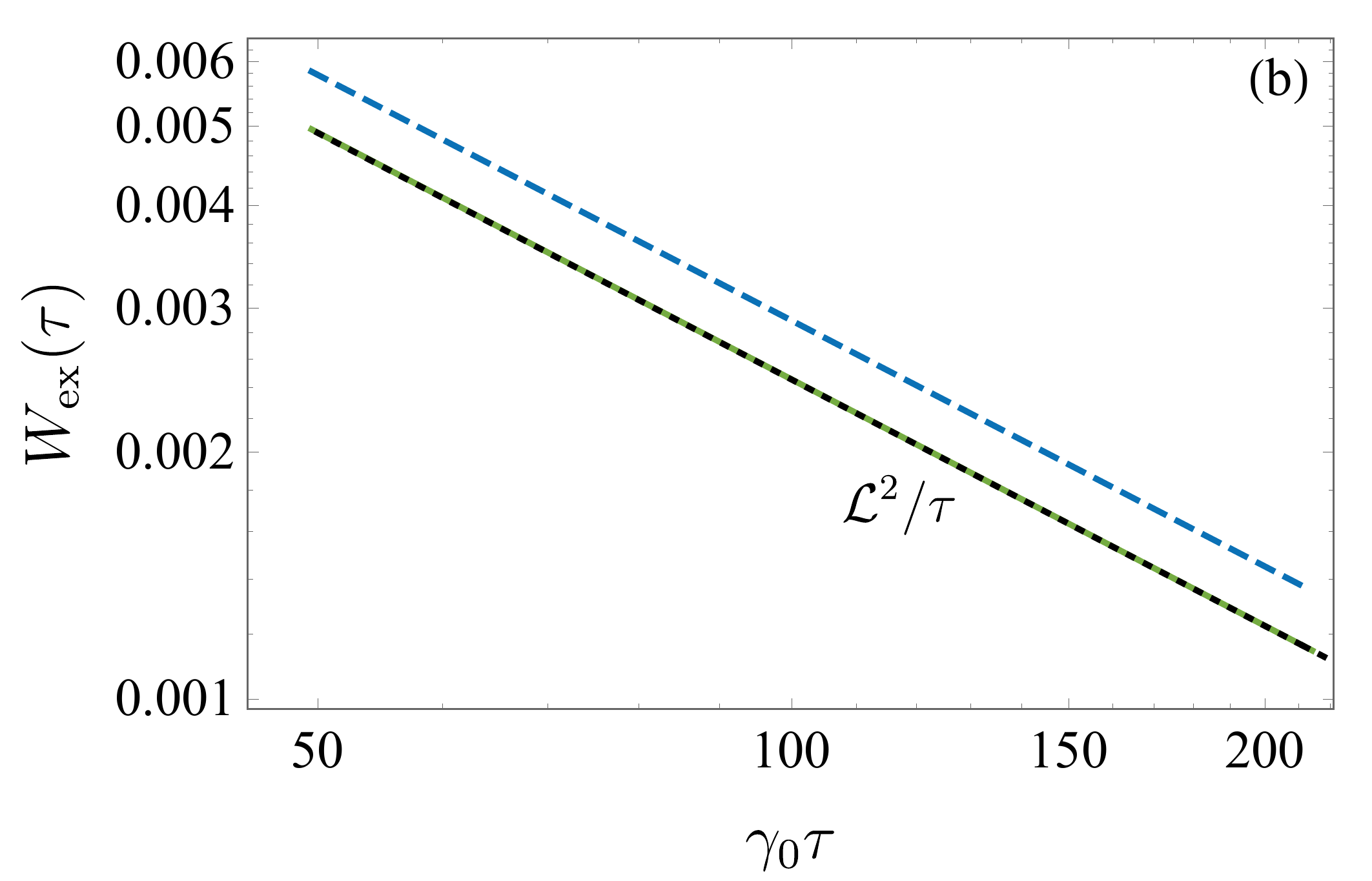}

\caption{Finite-time extrapolation for the two-level system with the tuned
energy spacing. (a) Finite-time thermodynamic lengths for the linear
and the optimal protocols. Through measuring the finite-time thermodynamic
length with short duration (markers), the thermodynamic length can
be approached by the extrapolated functions of both protocols. (b)
The $\tau^{-1}$ scaling of the excess work. For a long-time process,
the optimal protocol (green solid line) consumes less excess work,
and saturates the lower bound given by the thermodynamic length (black
dotted line). \label{fig:ontwolevelsystem}}

\end{figure}

With the Ohmic spectral $\alpha=1$ of the bath, we exemplify the
extrapolation method to measure the thermodynamic length of the two-level
system. The energy spacing is tuned from $E_{0}=1$ to $E_{\tau}=2$
with the linear and the optimal protocols. Figure \ref{fig:ontwolevelsystem}(a)
shows the measured finite-time thermodynamic length $\mathcal{L}(\tau)=\int_{0}^{\tau}\sqrt{P_{\mathrm{ex}}(t)}dt$
(markers) with the duration ranging from 10.0 to 19.0. With the cutoff
$N=3$, the extrapolated functions are obtained for the two protocols
as $\mathcal{L}_{\mathrm{li}}(\tau)=0.495-0.000284/\tau-0.0179/\tau^{2}-0.0307/\tau^{3}$
and $\mathcal{L}_{\mathrm{op}}(\tau)=0.495-0.0136/\tau-0.0500/\tau^{2}-0.0130/\tau^{3}$.
They both approach the thermodynamic length (black dotted line) with
the increasing duration $\tau$. Figure \ref{fig:ontwolevelsystem}(b)
shows the $\tau^{-1}$ scaling of the excess work at the long-time
limit. The excess work is bounded by the thermodynamic length as $W_{\mathrm{ex}}\geq\mathcal{L}^{2}/\tau$.
The excess work done with the optimal protocol matches the lower bound
given by the thermodynamic length (black dotted line).

\section{quantum Brownian motion in a tuned harmonic potential\label{sec:Case-study:-quantum}}

\subsection{Differential equations of the average values}

We first derive the differential equations of the average values of
the internal energy $\left\langle H(t)\right\rangle =\mathrm{Tr}[\rho(t)H(t)]$,
the Lagrange $\left\langle L(t)\right\rangle =\mathrm{Tr}[\rho(t)L(t)]$
and the correlation function $\left\langle D(t)\right\rangle =\mathrm{Tr}[\rho(t)D(t)]$
in the main content. According to the Caldeira-Leggett master equation
\citep{Caldeira1983,Breuer2007}, the time derivative of the internal
energy is calculated as 
\begin{equation}
\frac{d}{dt}\left\langle H\right\rangle =m\omega(t)\dot{\omega}(t)\mathrm{Tr}[\rho(t)\hat{x}^{2}]+\mathrm{Tr}\{\mathscr{L}_{\omega(t)}[\rho(t)]H(t)\},
\end{equation}
where the upper dot denotes the time derivative, namely $\dot{\omega}(t)=d\omega/dt$.
With the similar calculation of time derivatives for the Lagrange
and the correlation function, we obtain the evolution of average values
as

\begin{equation}
\frac{d}{dt}\left\langle H\right\rangle =\frac{\dot{\omega}}{\omega}\left(\left\langle H\right\rangle -\left\langle L\right\rangle \right)-2\kappa(\left\langle H\right\rangle +\left\langle L\right\rangle )+2\kappa k_{B}T_{\mathrm{b}},
\end{equation}

\begin{equation}
\frac{d}{dt}\left\langle L\right\rangle =-\frac{\dot{\omega}}{\omega}\left(\left\langle H\right\rangle -\left\langle L\right\rangle \right)-2\omega\left\langle D\right\rangle -2\kappa(\left\langle H\right\rangle +\left\langle L\right\rangle )+2\kappa k_{B}T_{\mathrm{b}},
\end{equation}
\begin{equation}
\frac{d}{dt}\left\langle D\right\rangle =\frac{\dot{\omega}}{\omega}\left\langle D\right\rangle +2\omega\left\langle L\right\rangle -2\kappa\left\langle D\right\rangle ,
\end{equation}
The three equations above are Eqs. (8) in the main context.

Let us rewrite these differential equations into a compact form

\begin{equation}
\frac{d}{dt}\phi=M(t)\phi+f(t),\label{eq:phi=00003DMphi+f}
\end{equation}
with the vector $\phi=\left(\begin{array}{ccc}
\left\langle H\right\rangle  & \left\langle L\right\rangle  & \left\langle D\right\rangle \end{array}\right)^{\mathrm{T}}$ and the function $f(t)=2\kappa k_{B}T_{\mathrm{b}}\left(\begin{array}{ccc}
1 & 1 & 0\end{array}\right)^{\mathrm{T}}$. The matrix $M(t)$ is

\begin{equation}
M(t)=\left(\begin{array}{ccc}
-2\kappa+\frac{\dot{\omega}}{\omega} & -2\kappa-\frac{\dot{\omega}}{\omega} & 0\\
-2\kappa-\frac{\dot{\omega}}{\omega} & -2\kappa+\frac{\dot{\omega}}{\omega} & -2\omega\\
0 & 2\omega & -2\kappa+\frac{\dot{\omega}}{\omega}
\end{array}\right),\label{eq:M(t)}
\end{equation}
the inverse of which is obtained as

\begin{equation}
M^{-1}=\left(\begin{array}{ccc}
-\frac{\left(\frac{\dot{\omega}}{\omega}-2\kappa\right)^{2}+4\omega^{2}}{4\left(\frac{\dot{\omega}}{\omega}-2\kappa\right)\left(2\kappa\frac{\dot{\omega}}{\omega}-\omega^{2}\right)} & \frac{\frac{\dot{\omega}}{\omega}+2\kappa}{4\left(\omega^{2}-2\kappa\frac{\dot{\omega}}{\omega}\right)} & \frac{\omega\left(\frac{\dot{\omega}}{\omega}+2\kappa\right)}{2\left(\frac{\dot{\omega}}{\omega}-2\kappa\right)\left(\omega^{2}-2\kappa\frac{\dot{\omega}}{\omega}\right)}\\
\frac{\frac{\dot{\omega}}{\omega}+2\kappa}{4\left(\omega^{2}-2\kappa\frac{\dot{\omega}}{\omega}\right)} & \frac{2\kappa-\frac{\dot{\omega}}{\omega}}{4(2\kappa\frac{\dot{\omega}}{\omega}-\omega^{2})} & \frac{\omega}{2\left(\omega^{2}-2\kappa\frac{\dot{\omega}}{\omega}\right)}\\
\frac{\omega\left(\frac{\dot{\omega}}{\omega}+2\kappa\right)}{2\left(\frac{\dot{\omega}}{\omega}-2\kappa\right)\left(2\kappa\frac{\dot{\omega}}{\omega}-\omega^{2}\right)} & \frac{\omega}{2(2\kappa\frac{\dot{\omega}}{\omega}-\omega^{2})} & \frac{\gamma\frac{\dot{\omega}}{\omega}}{4\left(\frac{\dot{\omega}}{\omega}-2\kappa\right)\left(2\kappa\frac{\dot{\omega}}{\omega}-\omega^{2}\right)}
\end{array}\right).
\end{equation}

\subsection{Solution of slow tuning}

With the existence of the inverse $M^{-1}$, we rewrite the differential
equation as 
\begin{equation}
\phi=-M^{-1}f+M^{-1}\frac{d}{dt}\phi.
\end{equation}
Using the perturbative expansion approach in Ref. \citep{Cavina2017},
the solution for the slow tuning is
\begin{equation}
\phi=-\sum_{n=0}^{\infty}(M^{-1}\frac{d}{dt})^{n}M^{-1}f,\label{eq:17}
\end{equation}
where the time derivative $d/dt$ acts on both $M^{-1}$ and $f$.
For the slow tuning satisfied 
\begin{equation}
\frac{\dot{\omega}}{\kappa\omega}\ll1,\:\frac{\kappa\dot{\omega}}{\omega^{3}}\ll1,
\end{equation}
the term with $n=0$ dominates the summation in Eq. (\ref{eq:17}),
namely 
\begin{equation}
\phi\approx-M^{-1}f=k_{B}T_{\mathrm{b}}\left(\begin{array}{c}
\frac{\kappa\left[2\omega^{2}+\frac{\dot{\omega}}{\omega}(\frac{\dot{\omega}}{\omega}-2\kappa)\right]}{(2\kappa-\frac{\dot{\omega}}{\omega})\left(\omega^{2}-2\kappa\frac{\dot{\omega}}{\omega}\right)}\\
\frac{\kappa\dot{\omega}}{2\kappa\dot{\omega}-\omega^{3}}\\
-\frac{2\kappa\dot{\omega}}{(2\kappa-\frac{\dot{\omega}}{\omega})\left(\omega^{2}-2\kappa\frac{\dot{\omega}}{\omega}\right)}
\end{array}\right).
\end{equation}
Keeping the first order of $\dot{\omega}/(\kappa\omega)$, $\kappa\dot{\omega}/\omega^{3}$
and $\dot{\omega}/\omega^{2}$, we obtain
\begin{equation}
\phi^{[1]}\approx k_{B}T_{\mathrm{b}}\left(\begin{array}{c}
1+\kappa\frac{\dot{\omega}}{\omega^{3}}+\frac{1}{2\kappa}\frac{\dot{\omega}}{\omega}\\
-\kappa\frac{\dot{\omega}}{\omega^{3}}\\
-\frac{\dot{\omega}}{\omega^{2}}
\end{array}\right).\label{eq:zeroorderphi}
\end{equation}
 The terms with the change of the frequency $\dot{\omega}$ contributes
to the $\tau^{-1}$ scaling of the excess work at the long-time limit.

\begin{figure}
\includegraphics[width=7cm]{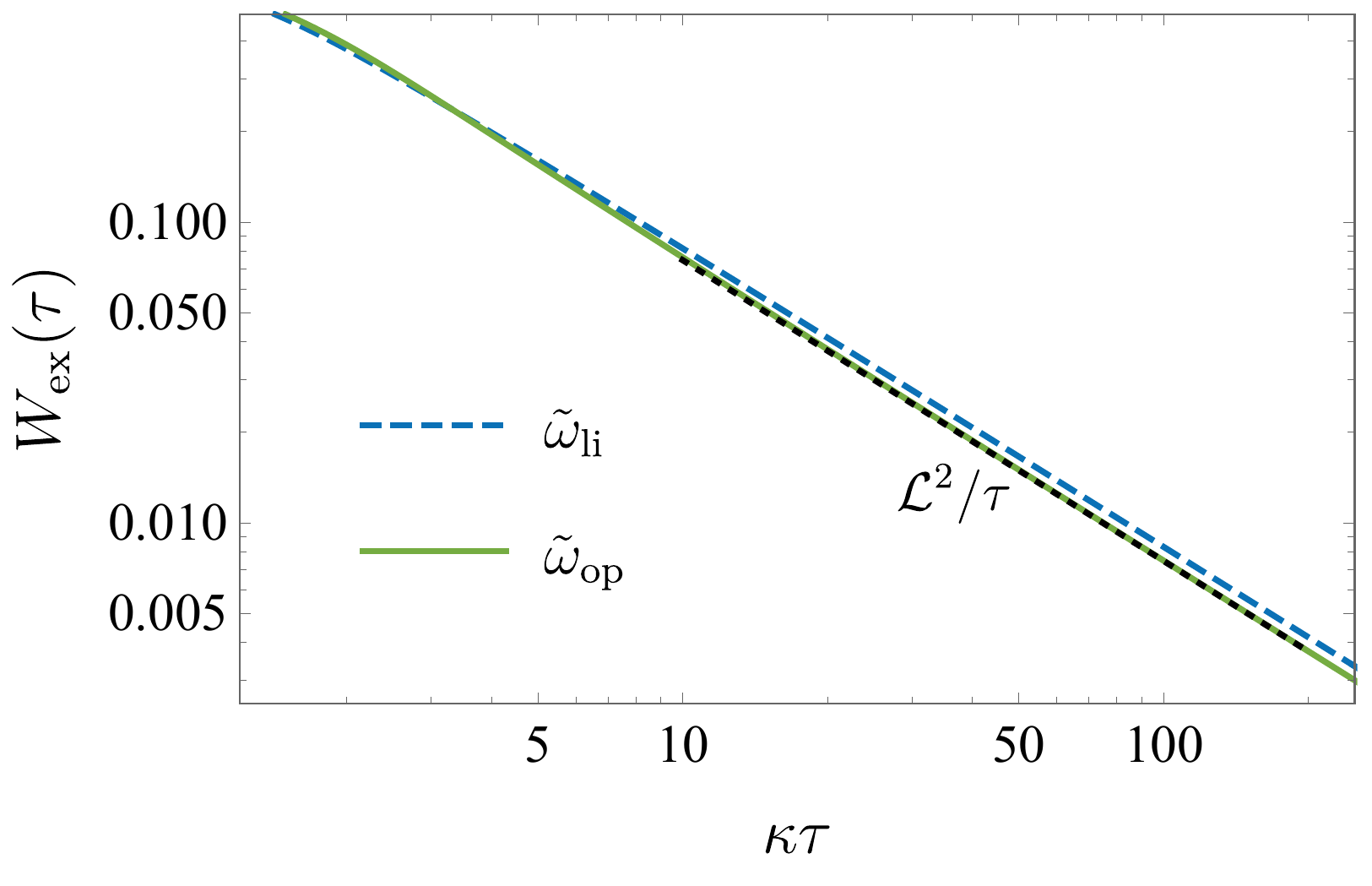}

\caption{$\tau^{-1}$ scaling of the excess work for tuning the frequency of
the harmonic oscillator. We compare the excess work in the linear
protocol (blue dashed curve) and the optimal protocol (green solid
curve). The frequency is tuned from $\omega_{0}=1$ to $\omega_{\tau}=2$
with the temperature $k_{B}T_{\mathrm{b}}=1$, and the dissipation
rate is set as $\kappa=1$. For the slow process, the lower bound
of the excess work $\mathcal{L}^{2}/\tau$ is given by the thermodynamic
length with $\mathcal{L}=0.864$. \label{fig:excesswork_harmonic_oscillator}}
\end{figure}

The power for tuning the frequency is 
\begin{align}
\dot{W}(t) & =\frac{\dot{\omega}}{\omega}\left(\left\langle H\right\rangle -\left\langle L\right\rangle \right),
\end{align}
and the quasi-static work rate is $\dot{W}_{(0)}(t)=k_{B}T_{\mathrm{b}}\dot{\omega}/\omega$.
Plugging Eq. (\ref{eq:zeroorderphi}) into the power, we obtain the
excess power $P_{\mathrm{ex}}(t)=\dot{W}(t)-\dot{W}_{(0)}(t)$ to
the lowest order of $\dot{\omega}$ as 
\begin{equation}
P_{\mathrm{ex}}(t)\approx k_{B}T_{\mathrm{b}}\frac{\dot{\omega}^{2}}{\omega^{2}}(\frac{1}{2\kappa}+\frac{2\kappa}{\omega^{2}}),
\end{equation}
which leads to the thermodynamic length

\begin{align}
\mathcal{L} & =\int_{\omega_{0}}^{\omega_{\tau}}\sqrt{k_{B}T_{\mathrm{b}}\left(\frac{1}{2\kappa}+\frac{2\kappa}{\omega^{2}}\right)}\left|\frac{d\omega}{\omega}\right|.
\end{align}

\subsection{Numerical result of the excess work}

In Fig. \ref{fig:excesswork_harmonic_oscillator}, we show the numerical
result of the excess work corresponding to the subset of Fig. 2 in
the main content. We consider the linear protocol (blue dashed curve)
with $\tilde{\omega}_{\mathrm{li}}(s)=\omega_{0}+(\omega_{\tau}-\omega_{0})s$
and the optimal protocol (green solid curve) with $\tilde{\omega}_{\mathrm{op}}(s)$
solved by Eq. (10) in the main content. For the slow tuning with long
duration, the excess work satisfies the $\tau^{-1}$ scaling, and
is bounded by the thermodynamic length as $W_{\mathrm{ex}}\geq\mathcal{L}^{2}/\tau$.
The lower bound (black dotted line) is saturated by the optimal protocol.

\section{Compression of classical ideal gas\label{sec:Case-study:-Compression}}

Our recent experimental setup for validation of the $\tau^{-1}$ scaling
of the excess work with the ideal gas \citep{Ma2019} can be used
to apply the current strategy to measure the thermodynamic length.
Here we show the theoretical derivation of the thermodynamic length
and the excess work for the finite-time compression.

\begin{figure}
\includegraphics[width=7cm]{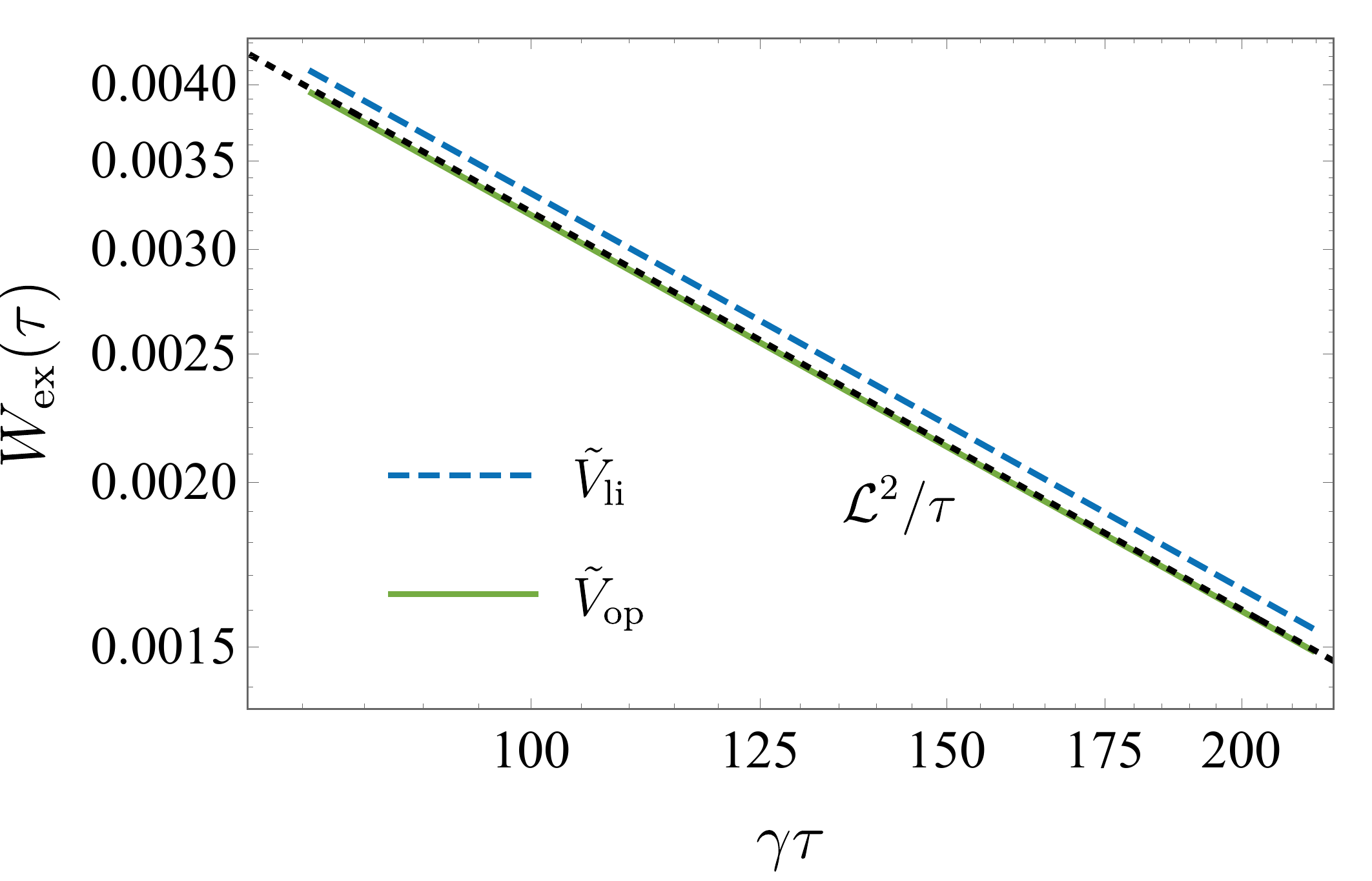}

\caption{$\tau^{-1}$ scaling of the excess work for the finite-time compression
of the ideal gas. The linear protocol $\tilde{V}_{\mathrm{li}}(s)=V_{0}+(V_{\tau}-V_{0})s$
and the exponential protocol $\tilde{V}_{\mathrm{op}}(s)=V_{0}(V_{\tau}/V_{0})^{s}$
are considered. The parameters are set as $nR=T_{\mathrm{b}}=\gamma=1$
and $C_{V}=1.5$ with the volume tuned from $V_{0}=1$ to $V_{\tau}=0.5$.
The black dotted line shows the lower bound of the excess work $W_{\mathrm{ex}}\protect\geq\mathcal{L}^{2}/\tau$
given by the thermodynamic length $\mathcal{L}=0.566$, saturated
by the exponential protocol. \label{fig:excess_work_finite_time_compression}}
\end{figure}

Defining the temperature difference $u=T-T_{\mathrm{b}}$, we rewrite
Eq. (11) in the main content as

\begin{equation}
\frac{du}{dt}=-\frac{nR}{C_{V}}(T_{\mathrm{b}}+u)\frac{\dot{V}}{V}-\gamma u.
\end{equation}
The series expansion solution is obtained as

\begin{align}
u & =\sum_{n=0}^{\infty}\left(-\frac{1}{\frac{nR}{C_{V}}\frac{\dot{V}}{V}+\gamma}\frac{d}{dt}\right)^{n}\left(\frac{-\frac{nRT_{\mathrm{b}}}{C_{V}}\frac{\dot{V}}{V}}{\frac{nR}{C_{V}}\frac{\dot{V}}{V}+\gamma}\right).\label{eq:ideal_gas_u}
\end{align}
The power of the compression is $\dot{W}=-p\dot{V}=-nRT\dot{V}/V$.
In the quasi-static process, the temperature of the gas is the same
as that of the bath, i.e. $T=T_{\mathrm{b}}$, and the quasi-static
work rate is $\dot{W}_{(0)}(t)=-nRT_{\mathrm{b}}\dot{V}/V$. For the
slow tuning $(nR/\gamma C_{V})(\dot{V}/V)\ll1$, the term with $n=0$
dominates the summation in Eq. (\ref{eq:ideal_gas_u}), and the temperature
difference to the lowest order is $u^{[1]}=-(nRT_{\mathrm{b}}/\gamma C_{V})(\dot{V}/V)$.
The excess power $P_{\mathrm{ex}}=\dot{W}-\dot{W}_{(0)}(t)$ approximates

\begin{align}
P_{\mathrm{ex}} & \approx\frac{(nR)^{2}T_{\mathrm{b}}}{\gamma C_{V}}\left(\frac{\dot{V}}{V}\right)^{2},
\end{align}
with the excess work
\begin{equation}
W_{\mathrm{ex}}\approx\int_{0}^{\tau}\frac{(nR)^{2}T_{\mathrm{b}}}{\gamma C_{V}}\left(\frac{\dot{V}}{V}\right)^{2}dt.
\end{equation}
The thermodynamic length follows as

\begin{align}
\mathcal{L} & =\int_{0}^{\tau}\sqrt{\frac{(nR)^{2}T_{\mathrm{b}}}{\gamma C_{V}}\left(\frac{\dot{V}}{V}\right)^{2}}dt,\label{eq:27}
\end{align}
which gives Eq. (12) in the main content.

In Fig. \ref{fig:excess_work_finite_time_compression}, we show the
numerical result of the excess work for the finite-time compression
of ideal gas. The excess work exhibits the $\tau^{-1}$ scaling during
the slow compression process. The compression with the exponential
protocol (green solid curve) consumes the lower excess work compared
to that of the linear protocol (blue dashed curve) with given duration
$\tau$. The lower bound $\mathcal{L}^{2}/\tau$ (black dotted line)
is saturated by the optimal protocol as the exponential protocol.

\bibliographystyle{apsrev4-1}
\bibliography{mainref}